\documentclass[sigconf]{acmart}

\AtBeginDocument{%
  \providecommand\BibTeX{{%
    \normalfont B\kern-0.5em{\scshape i\kern-0.25em b}\kern-0.8em\TeX}}}

\setcopyright{acmlicensed}
\copyrightyear{2018}
\acmYear{2018}
\acmDOI{XXXXXXX.XXXXXXX}

\acmConference[Conference acronym 'XX]{Make sure to enter the correct
  conference title from your rights confirmation emai}{June 03--05,
  2018}{Woodstock, NY}
\acmISBN{978-1-4503-XXXX-X/18/06}

\usepackage[utf8]{inputenc} 
\usepackage{hyperref}       
\usepackage{url}            
\usepackage{paralist}
\usepackage{nicefrac}       
\usepackage{microtype}      
\usepackage{enumitem}
\usepackage{graphicx}
\usepackage{pifont}
\usepackage{tablefootnote}
\usepackage{threeparttable,color}
\usepackage{booktabs}
\usepackage{multicol}
\usepackage{multirow}
\usepackage{xspace}
\usepackage{algorithmic}

\usepackage{textcomp}
\usepackage{xcolor}
\usepackage{adjustbox}
\usepackage{subcaption}
\usepackage{booktabs} 
\usepackage{array}
\usepackage{siunitx} 
\definecolor{deepgreen}{rgb}{0.0, 0.5, 0.0}
\usepackage{colortbl} 
\definecolor{Gray}{gray}{0.9}

\def\BibTeX{{\rm B\kern-.05em{\sc i\kern-.025em b}\kern-.08em
    T\kern-.1667em\lower.7ex\hbox{E}\kern-.125emX}}

\newcommand{\cmark}{\textcolor[rgb]{0,0.6,0}{\ding{51}}}%
\newcommand{\xmark}{\textcolor[rgb]{1,0,0}{\ding{55}}}%
\newcommand{\modelname}{\textit{SpaFormer}\xspace}

\begin{document}

\title{Single Cells Are Spatial Tokens: Transformers for Spatial Transcriptomic Data Denoising}

\author{%
Hongzhi Wen$^1$\,\,, Wenzhuo Tang$^1$\,\,, Wei Jin$^2$\,\,, Jiayuan Ding$^1$\,\,, Renming Liu $^1$\,\,, Xinnan Dai$^1$\,\,, \\Feng Shi$^3$\,\,, Lulu Shaing$^4$\,\,, Hui Liu$^1$\,\,, Yuying Xie$^1$}
\affiliation{%
\institution{$^1$Michigan State University, $^2$Emory University,
$^3$TigerGraph Company,
$^4$MD Anderson Cancer Center,\country{USA}}}
 
\email{{wenhongz,tangwen2,dingjia5, liurenmi,daixinna, liuhui7, xyy}@msu.edu}
\email{{wei.jin}@emory.edu,{lshang}@mdanderson.org}





  

\renewcommand{\shortauthors}{Wen, et al.}

\begin{abstract}
Spatially resolved transcriptomics brings exciting breakthroughs to single-cell analysis by providing physical locations along with gene expression. However, as a cost of the extremely high spatial resolution, the cellular level spatial transcriptomic data suffer significantly from missing values. While a standard solution is to perform imputation on the missing values, most existing methods either overlook spatial information or only incorporate localized spatial context without the ability to capture long-range spatial information. Using multi-head self-attention mechanisms and positional encoding, transformer models can readily grasp the relationship between tokens and encode location information. In this paper, by treating single cells as spatial tokens, we study how to leverage transformers to impute spatial transcriptomic data. In particular, investigate the following two key questions: (1) \textit{how to encode spatial information of cells in transformers}, and (2) \textit{ how to train a transformer for spatial transcriptomic imputation}. By answering these two questions, we present a transformer-based imputation framework, \modelname, for cellular-level spatial transcriptomic data. Extensive experiments demonstrate that \modelname outperforms existing state-of-the-art imputation algorithms on three large-scale datasets while maintaining superior computational efficiency.  

\end{abstract}



\maketitle

\section{Introduction} \label{sec:intro}
Spatial transcriptomic technologies have rapidly developed in recent years and emerged as next-generation tools for biomedical research. For instance, in-situ hybridization (ISH) based technology~\cite{lubeck2014single} produces detailed single-cell transcriptomic profiles along with the location of cells within a tissue, yielding deeper insights into cell identity and functionality than ever. However, as the number of profiled genes increases, the requirement for additional rounds of hybridization also increases, which elevates the potential for cumulative errors. As a result, ISH-based spatial transcriptomic data generally suffer from low mRNA capture efficiency and may miss a significant number of expressed genes, resulting in many false zero counts in observed gene expression data.  

An effective approach to mitigate this problem is applying imputation methods to rectify the false zeros. There are numerous types of imputation methods for {\it conventional transcriptomic data} (i.e., scRNA-seq data). Nevertheless, those methods tend to yield suboptimal performance when imputing spatial transcriptomic data, as they do not leverage the presented spatial information.
Notably, the spatial locations of cells provide important information about cell-cell interactions as well as cell similarities~\cite{wen2022bi}, which have great potential to advance the imputation process. For example, tumor cells usually show strong aggregation, and neighboring tumor cells with similar micro-environments in terms of ligands tend to have higher gene expression similarity than distant cells~\cite{browaeys2020nichenet}.

To effectively utilize spatial information, graph neural networks have been recently utilized for spatial transcriptomic analysis~\cite{li2022cell,wang2022sprod}. Concretely, graph neural networks (GNNs) are applied to the cell-cell neighboring graphs built on the spatial positions. \textbf{However, these methods are limited to modeling a localized spatial context, which can be unfavorable for identifying long-range correlated cells. }
For example, Treg cells are scarce spatially in many tissues but still tend to be homologous and share similar gene expression profiles~\cite{rudensky2011regulatory}. Hence, it is desirable to capture the cell interactions from broader contexts. To achieve this goal, we employ the transformer model~\cite{vaswani2017attention} for the studied problem. The transformer model was originally designed for textual data. It uses multi-head self-attention mechanisms to model relationships between input tokens and utilizes positional encoding to model the locations of tokens. Transformers are able to weigh the importance of each input token relative to all other tokens, rather than adjacent ones as in GNNs. In our studied problem, by treating cells as tokens, we can readily apply the transformer model to capture long-range correlations between cells. 

In this paper, we investigate two key questions when applying transformers to spatial transcriptomic imputation: (1) \textit{how to encode spatial information of cells in transformers}, and (2) \textit{how to train a transformer for transcriptomic imputation}. To answer the first question, one natural idea is to adopt the positional encodings from common transformers to encode spatial cellular coordinates. However, the spatial coordinates of cells are continuous and irregular, which are essentially different from the discrete coordinates in the common practice of transformers. Therefore, efforts are still desired to design positional encodings for spatial transcriptomics. To address this issue, we investigate existing positional encodings, compare their advantages and disadvantages, and conduct comprehensive experiments to obtain a best practice for spatial transcriptomics.
To answer the second question, we generalize the well-studied imputation models for conventional transcriptomic data into a flexible autoencoder framework, where we adopt a transformer as the encoder. Furthermore, we propose a new bi-level masking technique that can be plugged into the general autoencoder framework.  With the solutions to two questions, our proposed framework, \modelname, consistently achieves outstanding imputation performance on three large-scale cellular-level spatial transcriptomic datasets.

\section{Preliminary}

\subsection{Problem Statement} \label{sec:prob}

\begin{figure}[ht]%
     \centering
     \subfloat[Molecular image.]{\label{fig:ml}{\includegraphics[width=0.46\linewidth]{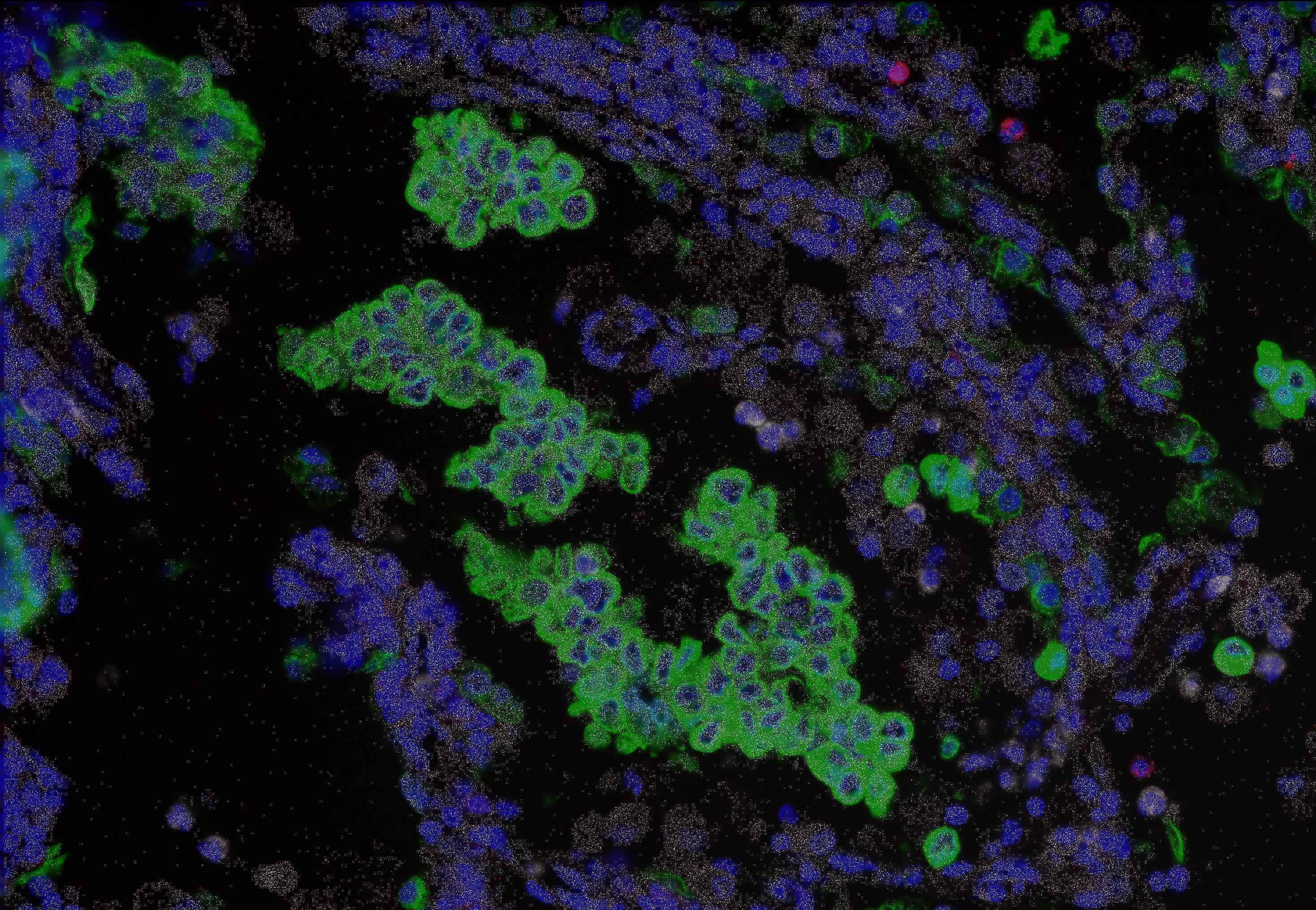} }}%
      \subfloat[Cell segmentation.]{\label{fig:cs}{\includegraphics[width=0.46\linewidth]{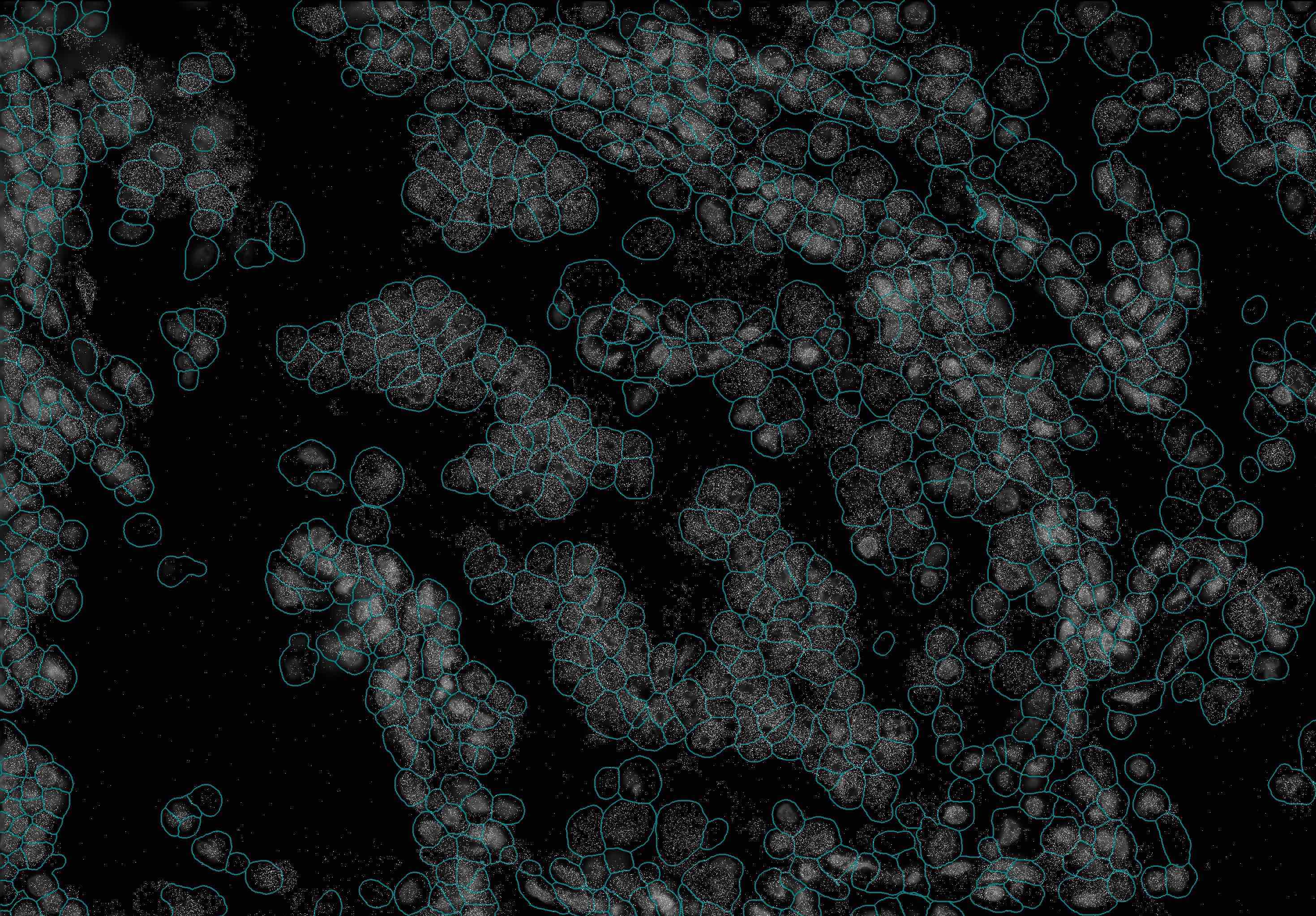} }}%
\qquad
\vskip -0.7em
\caption{A sample image of protein, RNA molecules, and segmented cells.  Colors in sub-figure (a) indicate the protein molecules that are stained. These proteins contribute to the cell segmentation process, which results in the sub-figure (b). The final output from the pipeline consists of the position of each cell and a cell-by-gene count matrix.}
\vskip -1em
\label{fig:datavis}
\end{figure}

Before we introduce the notations and basic concepts, we first introduce the data we use. It is important to note that the focus of this paper is on \textbf{high-resolution cellular-level datasets} (typically generated by ISH-based techniques), as opposed to the commonly studied spot-level datasets (e.g., 10X Visium, Seq-based data). In this work, we use two published and one unpublished dataset produced by the Nanostring CosMx~\cite{He2021.11.03.467020} platform. To obtain the cellular level gene expression, CellPose~\cite{stringer2021cellpose} software is applied to conduct cell segmentation. Figure~\ref{fig:datavis} gives an example of the raw image data and how cells are segmented.

After pre-processing, a typical single-cell spatial transcriptomic dataset is comprised of two essential components, i.e., the gene expression of cells and the corresponding spatial locations. We denote the gene expression data as a matrix $\mathbf{X} \in \mathcal{R}^{N \times k}$, where $N$ is the number of cells, and $k$ is the number of genes.  Hereby, $\mathbf{X}_{i,j}$ denotes the count of the $j$-th gene captured in the $i$-th cell. We use $\mathbf{C} \in \mathcal{R}^{N \times 2}$ to denote the two-dimensional coordinates of each cell, where those coordinates are based on the center position of each cell. 
Note that each dataset is composed of multiple field-of-views (FOVs). Each FOV contains thousands of cells and might not be adjacent to each other. Thus, we focus on units of FOVs by default.

In the spatial transcriptomic imputation problem, we suppose that a part of the input values in $\mathbf{X}$ are missing, denoted as a mask matrix $\mathbf{M} \in\{0,1\}^{N \times k}$, where the value of $\mathbf{X}_{i,j}$ can only be observed when $\mathbf{M}_{i,j}=1$. A partially observed data matrix $\mathbf{\widetilde{X}}$ is defined as:
\begin{equation}
    \mathbf{\widetilde{X}}_{i,j} = \left\{
    \begin{array}{ccc}
        0 & & \mathbf{M}_{i,j}=0\\
        \mathbf{X}_{i,j} & & \mathbf{M}_{i,j}=1\\
    \end{array}
    \right.
\end{equation}
Our objective is to predict the missing values $\mathbf{X}_{i,j}$ at $\mathbf{M}_{i, j} = 0$, given the partially observed data $\mathbf{\widetilde{X}}_{i,j}$ and the spatial positions $\mathbf{C}$. {\it Note that in this work, we treat cells as tokens thus we use these two terms exchangeably in the remainder of the paper.}

\vspace{-0.5em}
\subsection{Transformers}

An essential component of our \modelname framework is transformer~\cite{vaswani2017attention} encoders. The original transformer comes with an encoder-decoder architecture, while in this paper we only take advantage of a transformer encoder. Specifically, a transformer encoder consists of alternating layers of Multi-head Self-Attention (MSA) and Multi-Layer Perceptron (MLP) blocks. Prior to each block, Layer Normalization (LN) is performed, and a residual connection is established following every block. 
To be concrete, a transformer encoder can be formulated as:
\begin{align}
\mathbf{H}_0 & =\left[ \mathbf{x}_1 \mathbf{E} ; \mathbf{x}_2 \mathbf{E} ; \cdots ; \mathbf{x}_n \mathbf{E}\right]+\mathbf{E}_\mathrm{pos}, & & \mathbf{E} \in \mathbb{R}^{k \times d}, \mathbf{E}_\mathrm{pos} \in \mathbb{R}^{n \times d} \\
\label{eq:attn} \mathbf{H}_{\ell}^{\prime} & =\operatorname{MSA}\left(\operatorname{LN}\left(\mathbf{H}_{\ell-1}\right)\right)+\mathbf{H}_{\ell-1}, & & \ell=1 \ldots L \\
\label{eq:H} \mathbf{H}_{\ell} & =\operatorname{MLP}\left(\operatorname{LN}\left(\mathbf{H}_{\ell}^{\prime}\right)\right)+\mathbf{H}_{\ell}^{\prime}, & & \ell=1 \ldots L
\end{align}
where $\mathbf{x}_i$ is $i$-th input token, i.e., $i$-th cell in an FOV, $\mathbf{H}_\ell \in \mathcal{R}^{n \times d}$ is the embeddings of tokens in the $\ell$-th layer, $n$ is the number of the input cells, $d$ is the hidden dimension, $\mathbf{E}$ is a learnable transformation layer, $\mathbf{E}_\mathrm{pos}$ is positional encodings for each cell. In our cases, positional encodings mainly encode the spatial positions, which is detailed in Section~\ref{sec:pe}. 

In our implementation, we adopt Performer~\cite{choromanski2021rethinking}, an efficient variant of transformers. Performer proposed high-efficient kernelized attention to approximate the original attention in Eq.~\ref{eq:attn}. As a result, Performer reduces the time complexity of transformers from quadratic to linear w.r.t number of input tokens, so that they can be applied to thousands of cell tokens within each FOV. In the following sections, we will first introduce how to capture spatial information as positional embedding in transformers and then detail the proposed framework. 

\begin{table*}[] 
\centering
\caption{Comparison between positional encodings regarding the desired properties for spatial transcriptomics.} \label{tab:pe}
\vskip -1em
\begin{tabular}{@{}ccccccc@{}}
\toprule
             & Sources            & Distance Aware.     & Global Effec.      & Translation Invari. & Structure Aware.     & Other Issues   \\ \midrule
Sinusoid PE  & Patch / Coordinate & \cmark & \cmark & \xmark   & \xmark &                \\
Learnable PE & Patch              & \cmark & \cmark & \xmark   & \xmark &                \\
Naive PE     & Coordinate         & \cmark & \cmark & \xmark   & \xmark &                \\
RWPE         & Graph              & \xmark & \xmark & \cmark   & \cmark &                \\
LapPE        & Graph              & \cmark & \cmark & \cmark   & \cmark & Sign Ambiguity \\
Relative PE  & Distance           & \cmark & \cmark & \cmark   & \xmark & Scalability   \\ \midrule
SignNet~\cite{lim2022sign}      & Model              & \cmark & \cmark & \cmark   & \cmark &                \\
Cond PE      & Model              & \cmark & \xmark & \cmark   & \cmark &                \\ \bottomrule
\end{tabular}
\end{table*}

\section{Encoding Spatial Information in Transformers} \label{sec:pe}
When employing transformer models to spatial transcriptomic data, 
a critical challenge is how to encode spatial information in transformers.
In a standard transformer, positional encodings (PEs) are added to the token embeddings to make use of the positional information of tokens. Thus, it paves us a way to capture spatial transcriptomics by producing positional encoding for every single cell. However, the coordinates of spatial transcriptomics are continuous and irregular, which are essentially different from the sequential or image data. To tackle this issue, we investigate three general groups of positional encodings and demonstrate how we apply them to spatial transcriptomics. On top of that, we introduce two advanced model-based positional encodings that address limitation of previous positional encodings. 

\noindent\textbf{Patch-based Positional Encodings.}
We define patch-based positional encodings as encodings derived from discrete and regular patches. For example, ViT~\cite{dosovitskiy2020image} separates an image into $16\times16$ patches, so that patches lie in a regular grid. Positional encodings generated from these patch coordinates are considered patch-based positional encodings. 
To apply this approach to spatial transcriptomic data, we segment a whole input region, a.k.a, a field-of-view (FOV), into regular patches. 
In this paper, we
segment each FOV into $100\times100$ patches, where each patch contains $0.3$-$0.55$ cell on average due to the distinct cell size of different tissues. Based on the patch coordinates, there are generally two ways to produce positional encodings (PE) for each patch, i.e., \textbf{\textit{learnable PE}} and \textbf{\textit{sinusoid PE}}. \textit{Learnable PE} is introduced in ViT~\cite{dosovitskiy2020image}, where it learns a positional encoding for each individual position. 
\textit{Sinusoid PE} is proposed in vanilla transformer~\cite{vaswani2017attention}, originally designed for sequential data, while it can be extended to 2-dimensional space~\cite{carion2020end}. 

\noindent\textbf{Coordinate-based Positional Encodings.}
Coordinate-based positional encodings aim to generate positional encodings directly from the continuous spatial coordinates. In order to improve the generalizability, we normalize the spatial coordinates $\mathbf{C} \in \mathcal{R}^{n \times 2}$ to $[0, 1]$ by applying min-max normalization in each FOV. 
In this subsection, we discuss two specific types of positional encodings: \textbf{\textit{naive PE}} and \textbf{\textit{sinusoid PE}}. \textit{naive PE} projects normalized spatial coordinates to desired dimension $d$ via an MLP or other transformation. Compared to patch-based \textit{learnable PE}, \textit{naive PE} imposes the ordinal relation between coordinates. On the other hand, coordinate-based \textit{sinusoid PE} replaces discrete patch coordinates in the patch-based version with continuous spatial coordinates.

\noindent\textbf{Graph-based Positional Encodings.}
Graph-based positional encodings are derived from the spatial adjacency graph that connects adjacent tokens. 
The motivation of graph-based positional encodings is that the spatial adjacency graph conveys the relative position relations between tokens so that we can capture the positional information by encoding the spatial adjacency graph. To this end, we construct a spatial graph in which cell pairs are connected when the euclidean distance is smaller than $15$-$25 \mu m$. As a result, each cell is connected to $4$-$6$ cells on average, depending on the specific tissue type. The resulting adjacency matrix is denoted as $\mathbf{A}$. Next, we present two ways to encode positional information from spatial adjacency graphs:
\textbf{\textit{random walk PE}} (RWPE) , which derives from landing probabilities of random walks, is proposed in LSPE~\cite{dwivedi2021graph}. 
This positional encoding is effective in encoding graph structures, however, it hardly encodes distance information. Furthermore, it is limited to structural information within the $k$-th order neighborhood and thus overlooks the global context.
\textbf{\textit{Laplacian PE}} (LapPE), which uses Laplacian eigenvectors as positional encodings~\cite{kreuzer2021rethinking}. Specifically, graph Laplacian $\mathbf{L}$ is defined as $\mathbf{L}=\mathbf{I}_n-\mathbf{D}^{-1 / 2} \mathbf{A} \mathbf{D}^{-1 / 2}=\mathbf{U} \boldsymbol{\Lambda} \mathbf{U}^T$, where $\mathbf{I}_n$ is an identity matrix, and $\boldsymbol{\Lambda} \in \mathcal{R}^{k \times k}$ and $\mathbf{U} \in \mathcal{R}^{n \times k}$ correspond to the eigenvalues and eigenvectors respectively, $n$ is the number of nodes, and $k$ is the number of top eigenvalues we select. The Laplacian eigenvectors constitute a local coordinate system that retains the overall structure of the graph. Thus, we can use the $i$-th row of eigenvector matrix $\mathbf{U}$ as the positional encoding for node~$i$ (a.k.a, cell~$i$) in the graph. LapPE is generally a good choice despite being limited by sign ambiguity~\cite{kreuzer2021rethinking}. A straightforward approach to address this problem is to randomly flip the sign of eigenvectors during training, to force the model to be sign-invariant.

\begin{figure*}[ht]
    \vskip -1em
     \centering
     \label{fig:x}{\includegraphics[width=0.9\linewidth]{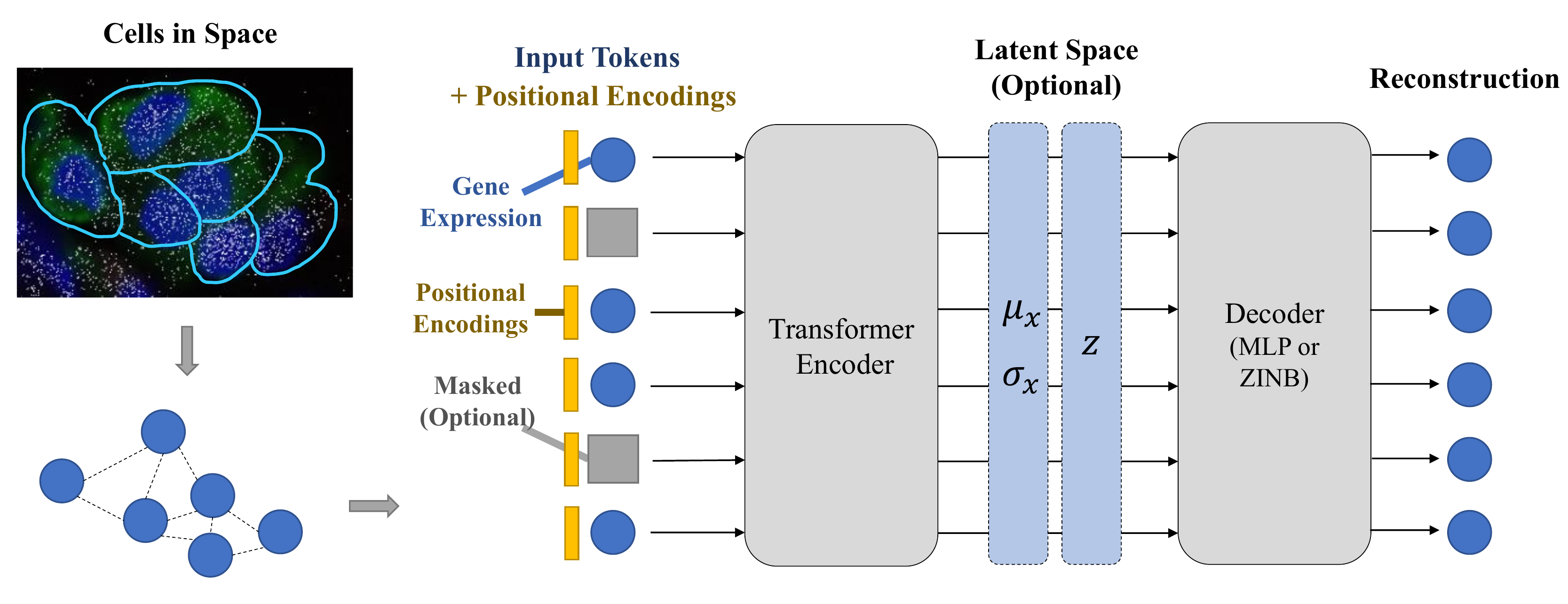} }
\vspace{-1em}
\caption{An illustration of our transformer-based autoencoder framework for spatial transcriptomics data imputation.} \label{fig:frame}
\label{fig:abl}%
\vspace{-0.5em}
\end{figure*}

\noindent\textbf{Model-based Positional Encodings.}
Despite that the aforementioned positional encodings can encode spatial information, each of them has certain limitations. Specifically, patch-based and coordinate-based positional encodings are dependent on the absolute location. However, studies have shown that relative positional encodings~\cite{wu2021rethinking, luo2021stable}, which consider the pair-wise relationships, generally perform better in various domains. One key advantage of relative positional encodings is their translation-invariant property. Translation-invariance refers to the property that positional encodings remain unchanged upon global translation of the coordinates, thereby enhancing the generalizability of transformers. 
Fortunately, graph-based positional encodings also achieve translation invariance since they are derived from the spatial adjacency graph, which is naturally based on invariant pair-wise distances. Nevertheless, RWPE cannot accurately encode pair-wise distance, while LapPE suffers from sign ambiguity. 

To overcome the aforementioned limitations, we introduce two advanced models to generate positional encodings based on spatial adjacency graphs: 
\textit{\textbf{SignNet}}~\cite{lim2022sign} proposes a sign invariant and permutation invariant network to learn positional encodings from the Laplacian eigenvectors of a target graph. The resulting positional encodings inherit the advantages of LapPE while not suffering from sign ambiguity.
\textit{\textbf{Cond PE}}, which stands for conditional positional encodings, originally proposed in CPVT~\cite{chu2021conditional}. In the original version, CPVT added a simple convolution layer before vision transformers, to provide positions conditioned on the local neighborhood of each input token. We adopt this idea while substituting the visual convolution with a graph convolution, as it is more feasible to implement the graph convolution on a spatial graph compared to directly convolute the sparsely positioned cell tokens in 2-dimensional space. Consequently, we apply a graph attention network~\cite{velivckovic2017graph} to the spatial adjacency graph to generate initial embedding for each cell token. This embedding is considered a conditional positional encoding (Cond PE) that encodes spatial neighborhoods while achieving translation invariance. 

\noindent\textbf{Summary.}
We summarize all aforementioned positional encodings in Table~\ref{tab:pe}. We consider four main properties when we compare these methods, namely distance awareness, global effectiveness, translation invariance, and spatial structure awareness. These properties have already been mentioned in the previous introduction, and more details can be found in Appendix~\ref{app:pe}. In conclusion, we investigate four types of position encoding to capture spatial information in transformers, which possess distinct properties and mostly demonstrate good potential as position encodings for spatial transcriptomics. We will evaluate all these positional encodings in our \modelname framework in order to gain the best practices for position encoding in spatial transcriptomes.


\vspace{-0.1em}
\section{Our Framework: \modelname}
\vspace{-0.2em}
In this section, we introduce our \modelname framework. An overview of \modelname is illustrated in Figure~\ref{fig:frame}. In \modelname, we first extract the spatial positional encoding for each cell, using different methods as discussed in Section~\ref{sec:pe}. Then cell embeddings are initialized with gene expressions and positional encodings, while some cells are selectively masked depending on the specific setting. Next, a transformer encoder is applied to encode both cellular profiles and intercellular contexts into the latent space.
Finally, a decoder reconstructs the input (or masked) information based on the latent variables. In the following, we first introduce our general framework that generalizes popular transcriptomic imputation models in Section~\ref{sec:autoencoder}. Then based on the general framework, we propose a new bi-level masking strategy in Section~\ref{sec:mae}, which is particularly suitable for spatial transcriptomic data imputation.

\vspace{-0.4em}
\subsection{Generalized Autoencoder Framework} \label{sec:autoencoder}
\vspace{-0.1em}
The most popular architecture for deep-learning-based transcriptomic data imputation methods~\cite{molho2022deep} is autoencoder, due to its prevalence in data denoising and missing data applications. 
Existing methods adopt a few variants of autoencoders, including variational autoencoders (VAE)~\cite{kingma2013auto, lopez2018deep} and ZINB-based (zero-inflated negative binomial) autoencoders~\cite{eraslan2019single}, while they often lack a systematic comparison for these autoencoder variants. In order to compare the performances of different autoencoders on the spatial transcriptomic imputation task, our \modelname framework generalizes all these variants of autoencoders. 

In the general framework of \modelname, our model takes a batch of cells as input. To be consistent with Section~\ref{sec:prob}, we denote the input as $\mathbf{\widetilde{X}} \in \mathcal{R}^{n \times k}$, where $n$ is the number of cells in the input field-of-view (FOV). Note that here we omit the positional encodings, which should also be included in the input matrix. An encoder $q_\theta$ projects the input data to latent space, resulting in $\mathbf{Z} \in \mathcal{R}^{n \times d}$, where $d$ is the latent dimension. A decoder $p_\phi$ then generates an output $\mathbf{\widehat{X}} \in \mathcal{R}^{n \times k}$ from the latent space, where we expect $\mathbf{\widehat{X}}$ to be identical with the input matrix $\mathbf{\widetilde{X}}$. The overall framework can be simply written as: 
\vspace{-1.5em}

\begin{align} \label{eq:encoder}
    \mathbf{Z} &= q_\theta\left(\mathbf{\widetilde{X}}\right)\\
    \mathbf{\widehat{X}} &= p_\phi\left(\mathbf{Z}\right)
\end{align}
\vskip -0.5em

In a vanilla autoencoder, $q_\theta$ and $p_\phi$ can be implemented by a deep learning model, such as a multi-layer perception (MLP) or transformer. We hereby suggest an asymmetrical encoder-decoder architecture (i.e., transformer as encoder $q_\theta$, MLP as decoder $p_\phi$), as we believe that the model requires a greater ability to utilize spatial information for denoising during the encoding process. During decoding, the contextual information of each cell has already been included in the latent space, making decoding easier. Next, we proceed to upgrade the model architecture with more advanced designs.

\noindent\textbf{ZINB-based autoencoders.} Previous studies~\cite{eraslan2019single, lopez2018deep} pointed out that the data distribution of the transcriptomic data can be approximated by zero-inflated negative binomial (ZINB) distribution because the data are discrete, overdispersed and contain many zero values. Therefore, we can adopt a ZINB-based autoencoder to leverage this prior information. A ZINB distribution is defined as:
\begin{equation}
\mathrm{NB}(x \mid \mu, \theta)=\frac{\Gamma(x+\theta)}{x ! \Gamma(\theta)}\left(\frac{\theta}{\theta+\mu}\right)^\theta\left(\frac{\mu}{\theta+\mu}\right)^x
\end{equation}
\begin{equation} \label{eq:zinb}
\operatorname{ZINB}\left(x \mid \pi, \mu, \theta\right)=\pi \delta_0(x)+(1-\pi) \operatorname{NB}(x)
\end{equation}
where $\mu$ and $\pi$ denote mean and dispersion, $\pi$ is the weight of the point mass at zero, and $\delta_0$ generates constant $0$.  

In our \modelname framework, to adapt a vanilla autoencoder to a ZINB-based autoencoder, the encoder $p_\phi$ remains unchanged, while a ZINB decoder implements the decoder $q_\theta$ instead of an MLP decoder. A ZINB decoder takes the latent code $\mathbf{Z}$ as input and generates an intermediate result $\widehat{\mathbf{H}} \in \mathcal{R}^{n \times d^{\prime}}$ via MLP. On top of that, ZINB decoders have three fully-connected output layers to estimate the parameters of ZINB distribution.
Let $\mathbf{\Pi}$, $\mathbf{M}$ and $\mathbf{\Theta}$ be the three parameters $\pi, \mu, \theta$ of ZINB distribution estimated by the three output layers respectively, then the overall loss function of ZINB-based autoencoder changes to the negative log-likelihood of the ZINB distribution, formulated as:
\begin{equation}
    \mathcal{L}_\textrm{ZINB} = -\operatorname{log}\left(\operatorname{ZINB}(\widetilde{\mathbf{X}} \mid \mathbf{\Pi}, \mathbf{M}, \mathbf{\Theta})\right)
\end{equation}
In the inference stage, the estimated mean matrix $\mathbf{M}$ is selected as the imputation output.

\noindent\textbf{Variational autoencoders.} VAEs have been wildly applied in single-cell transcriptomic analysis~\cite{lopez2018deep, lopez2019joint, molho2022deep} since they are robust to technical noise and bias. In our \modelname framework, we optionally transform an autoencoder framework to a VAE by making modifications to the latent space and loss function. Specifically, two additional fully-connected layers are appended to the encoder to estimate the mean and variance of latent variables:
\begin{align}
\boldsymbol{\mu}_z=q_\theta\left(\mathbf{\widetilde{X}}\right) \cdot \mathbf{W}_{\mu_z}, \quad
    \boldsymbol{\sigma}_z=\exp \left(q_\theta\left(\mathbf{\widetilde{X}}\right) \cdot \mathbf{W}_{\sigma_z} \right)
\end{align}
where $\mathbf{\mu_z}$ and  $\mathbf{\sigma_z}$ are mean and variance respectively. The latent variables $\mathbf{Z}$ are then sampled from the estimated Gaussian distribution $\mathcal{N}\left(\boldsymbol{\mu}_z, \boldsymbol{\sigma}_z\right)$, where we apply reparametrization trick~\cite{kingma2013auto} to keep the gradient descend possible. Furthermore, an additional Kullback-Leibler (KL) divergence loss term is added to regularize the distribution of latent variables, which is:
\begin{equation}
    \mathcal{L}_\mathrm{KL} = D_\mathrm{KL}\left(\mathcal{N}\left(\boldsymbol{\mu}_z, \boldsymbol{\sigma}_z\right) \Vert \mathcal{N}\left(0, 1\right)\right)
\end{equation}
The loss functions of vanilla autoencoder and ZINB-based autoencoder then turn to:
$\mathcal{L} = \mathcal{L}_\mathrm{MSE} + \beta \mathcal{L}_\mathrm{KL}$, or $\mathcal{L} = \mathcal{L}_\mathrm{ZINB} + \beta \mathcal{L}_\mathrm{KL}$, respectively, where $\beta$ is the KL loss weight. In the inference stage, we directly employ the estimated mean $\mathbf{\mu_z}$ as latent variables, instead of sampling.

\subsection{Bi-level Masked Autoencoders} \label{sec:mae}

Inspired by the tremendous success of the masked autoencoding paradigm in NLP~\cite{devlin2018bert} and Computer Vision~\cite{he2022masked}, we propose to incorporate a new masked autoencoder model into our generalized framework. Specifically, masked autoencoders are a form of more general denoising autoencoders~\cite{vincent2008extracting}, which adopt a simple concept to remove a proportion of the data and then learn to recover the removed parts. 
The idea of masked autoencoders is also natural and applicable in spatial transcriptomics since we expect a powerful model to be able to recover missing cellular profiles from neighboring cells sharing the same microenvironment and homologous cells with the same cell state. Note that the objective of the masked autoencoder is highly consistent with the process of data imputation. Hence, it is very promising that the model will be trained with the specific capability for imputing missing values.

Despite that masked autoencoders are highly suitable for training an imputation model, the characteristic of spatial transcriptomics adds difficulties to the utilization of masked autoencoders. Compared to NLP and vision, it is much more difficult to recover cell profiles solely based on contexts. One reason is that the external signals a cell receives depend not only on the concentration of ligands in the microenvironment but also on the amount of receptors on its membrane. Therefore, it's hardly possible to identify intercellular correlations when the cell profile is completely masked out. To address this issue, we made a modification to the standard masked autoencoder, which we call it a bi-level masking strategy.

In a bi-level masking strategy, for each input batch of tokens, we first determine which cells are to be masked. Then, instead of thoroughly masking the tokens, we selectively mask out a certain proportion of features in those tokens, leaving the potential for the model to recover underlying intercellular correlations. Specifically, we first sample a mask vector $\mathbf{\widetilde{m}}^\mathrm{token} \in \{0, 1\}^{n}$ from a Bernoulli distribution with $p=\theta$, where  $n$ is the number of input tokens, $\theta$ is the probability that a token is selected to be masked.
Next, $\mathbf{\widetilde{M}}^\mathrm{feat} \in \{0, 1\}^{n \times k}$ is sampled from another Bernoulli distribution with $p=\gamma$, where $\gamma$ is the probability that a feature is masked in a selected token. Lastly, we combined $\mathbf{\widetilde{m}}^\mathrm{token}$ and $\mathbf{\widetilde{M}}^\mathrm{feat}$  as:
\begin{equation}
    \mathbf{\widetilde{M}}_{i,j} = 1 - \mathbf{\widetilde{m}}^\mathrm{token}_{i} \cdot \mathbf{\widetilde{M}}^\mathrm{feat}_{i,j}  \label{eq:msk}
\end{equation}
where $\mathbf{\widetilde{M}}$ is the finalized mask matrix. The input features $\mathbf{\widetilde{X}}_{i, j}$ are then masked when $\mathbf{\widetilde{M}}_{i,j}=0$, resulting in a new feature matrix $\mathbf{\widetilde{X}}^\prime \in \mathcal{R}^{n \times k}$ written as:
\begin{equation}
    \mathbf{\widetilde{X}}^\prime = \phi\left(\widetilde{\mathbf{M}} \odot \mathbf{\widetilde{X}} \right)
\end{equation}
where $\odot$ indicates element-wise multiplication, and $\phi$ refers to a rescaling technique that maintains the mean of the input unchanged for each cell. Notably, when $\gamma=1$, our masked autoencoder is equivalent to MAE for vision~\cite{he2022masked}. When $\theta=1$, our masked autoencoder is equivalent to a denoising autoencoder~\cite{vincent2008extracting}. Such a bi-level masking strategy provides our framework with greater flexibility.

During training, masked features $\mathbf{\widetilde{X}}^\prime$ are used as initial token embeddings. Therefore, it should be plugged into Eq.~\ref{eq:encoder} to replace $\mathbf{\widetilde{X}}$. 
Note that our bi-level masking strategy perfectly fits the overall autoencoder framework, which means we can combine the masked autoencoder with other components, i.e., VAE and ZINB decoders. When enabling the masked autoencoder, we need to change the reconstruction loss in Eq.~\ref{loss:rec} to:
\begin{equation}
    \mathcal{L}_\mathrm{MSE} = \left\| \left(1-\mathbf{M}^{\prime}\right) \odot \left(\widehat{\mathbf{X}} - \mathbf{\widetilde{X}}\right) \right\|_F^2
\end{equation}
where we only calculate mean squared error (MSE) for the prediction of masked values.
For inference, we simply dismiss the bi-level masking strategy. Instead, the unmasked input $\mathbf{\widetilde{X}}$ is enabled, which provides extra information for imputation.

\begin{table*}[ht]
\centering
\caption{Imputation performance on three CosMx datasets. } \label{tab:main}
\vskip -1em
\begin{threeparttable}
\begin{tabular}{@{}ccccccccccc@{}}
\toprule
    &  \multirow{2}{*}{\textbf{Methods}} & \multicolumn{3}{c}{\textbf{Lung 5}}        & \multicolumn{3}{c}{\textbf{Kidney 1139}}   & \multicolumn{3}{c}{\textbf{Liver Normal}}            \\ 
        &    & RMSE$\downarrow{}{}$        & Pearson$\uparrow{}{}$      & Cosine$\uparrow{}{}$     & RMSE$\downarrow{}{}$        & Pearson$\uparrow{}{}$      & Cosine$\uparrow{}{}$    & RMSE$\downarrow{}{}$           & Pearson$\uparrow{}{}$         & Cosine$\uparrow{}{}$        \\ \midrule
Baseline & Raw     &    0.3758     &     -   &     -   &     0.3747     &    -      &     -   &     0.3507      &       -       &        -     \\ \midrule
\multirow{6}{*}{ \parbox{1.5cm}{\centering scRNA-seq \newline Methods}} &  scImpute     & 0.3245      & 0.444       & 0.5214     & 0.311       & 0.4824       & 0.5714    & 0.3048         & 0.4437          & 0.5074        \\
    & SAVER        & 0.3213      & 0.4564      & 0.5269     & 0.3106      & 0.4887       & 0.5689    & 0.2909         & 0.5462          & 0.5864        \\
    & scVI         & 0.2861      & \underline{0.6231}      & \underline{0.6661}     & 0.2901      & 0.5834       & 0.648     & 0.2797         & 0.5749          & 0.6224        \\
    & DCA          & \underline{0.2858}      & 0.6223      & 0.6648     & \underline{0.2852}      & \underline{0.5985}       & \underline{0.6597}    & 0.2542         & 0.657           & 0.688         \\
    & GraphSCI     & 0.3957      & 0.1334      & 0.3081     & 0.3624      & 0.2403       & 0.4128    & 0.3347         & 0.3707          & 0.443         \\
    & scGNN     & \multicolumn{3}{c}{OOM~\footnote{}} & \multicolumn{3}{c}{OOM~\textsuperscript{1}} & \multicolumn{3}{c}{OOM~\textsuperscript{1}} \\ \midrule

\multirow{3}{*}{ \parbox{1.5cm}{\centering Reference-based \newline Methods }} &  gimVI        & 0.3170       & 0.5320       & 0.5917     & 0.4387      & -0.0104      & 0.1967    & 0.4542         & -0.0015         & 0.1170         \\

    &  Tangram        &   0.3905     &  0.3161   &  0.3883   & 0.3639         & 0.4037         & 0.4798    & 0.3373      & 0.4830      & 0.5236  \\
    
    &  SpaGE        &  0.4420       & 0.3942       & 0.4572     & 0.5096      & 0.4311      & 0.5065    & 0.6433         & 0.5483         & 0.5899         \\ \midrule

\multirow{2}{*}{ \parbox{1.5cm}{\centering Spatial \newline Methods }} & Sprod        & \multicolumn{3}{c}{OOT~\footnote{}}               & \multicolumn{3}{c}{OOT\textsuperscript{2}}               & \multicolumn{3}{c}{OOT\textsuperscript{2}}            \\ 

    &  SEDR       & 0.3245  & 0.4949  & 0.5499           & 0.3116  & 0.5101  & 0.5826             & \multicolumn{3}{c}{OOM\textsuperscript{1}}            \\ \midrule

\multirow{2}{*}{Ours} & SpaGAT       &    0.2865	& 0.6047 &	0.6518     & 0.2859	 & 0.5852 &	0.6506        &    \underline{0.2241} &	\underline{0.7624}  &	\underline{0.7829}    \\
    &   SpaFormer & \bf0.2785     & \bf0.6363      & \bf0.6786     & \bf0.2794      & \bf0.6108       & \bf0.671     & \bf0.2117	& \bf0.7793 & \bf0.7973 \\ \bottomrule
\end{tabular}
\begin{tablenotes}
   \item[1] Run out of 300G CPU Memory. $\quad\quad\quad$ $^2$ Run more than 48 hours on 128 CPUs.
  \end{tablenotes}
\end{threeparttable}
\vskip -0.5em
\end{table*}

\section{Experiment}

\subsection{Experimental settings}
\textbf{Datasets.}
We validate the proposed approach on three spatial transcriptomic datasets generated by the CosMX platform~\cite{He2021.11.03.467020} from Nanostring, and can be accessed on their official website~\footnote{Dataset can be downloaded from \url{https://nanostring.com/products/cosmx-spatial-molecular-imager/nsclc-ffpe-dataset/}}. These datasets differ in their scales and sources of tissues, which highlights the comprehensiveness of our experiments. The data statistics are presented in Table~\ref{tab:data}.
\begin{table}[tb] 
\centering
\caption{Dataset statistics.}\label{tab:data}
\vspace{-1em}
\begin{tabular}{c|cccc}
\toprule
   Dataset       & Cell Num. & FOV Num. & Gene Num. & Zero Ratio\\ \midrule
Lung 5    & 99,656 & 30  & 960       & 86.74\%  \\ 
Kidney 1139 & 61,283 & 18 & 960       & 83.49\%    \\ 
Liver Normal & 305,730 & 244 & 1000       & 86.39\%  \\ \bottomrule
\end{tabular}
\vspace{-1em}
\end{table}

\noindent
\textbf{Baselines.}
To evaluate the effectiveness of \modelname{}, we compare it with the state-of-the-art spatial and non-spatial transcriptomic imputation models: {scImpute}~\cite{li2018accurate}, {SAVER}~\cite{huang2018saver}, {scVI}~\cite{lopez2018deep}, {DCA}~\cite{eraslan2019single}, {GraphSCI}~\cite{rao2021imputing}, {scGNN}~\cite{wang2021scgnn}, 
{gimVI}~\cite{lopez2019joint}, {Tangram}~\cite{biancalani2021deep}, {SpaGE}~\cite{abdelaal2020spage}, {Sprod}~\cite{li2022cell}, and {SEDR}~\cite{xu2024unsupervised}. The first six methods are designed for scRNA-seq data imputation. gimVI, Tangram, and SpaGE aim to impute spatial transcriptomic data via external reference scRNA-seq data. Sprod and SEDR build cell-cell graphs among spatial neighbors to help imputation. In addition to these published baselines, we create a new baseline model SpaGAT which uses the same bi-level masking autoencoder framework as \modelname, based on a graph neural network encoder with spatial graphs. Specifically, we implement a graph attention network~\cite{velivckovic2017graph} as the encoder. Since the graph attention network is a localized version of transformers, SpaGAT can be considered an ablation study for our \modelname model. 

\noindent
\textbf{Implementation Settings.}\label{sec:dp}
Before we conduct the experiment, we first randomly mask $30\%$ of the data to create the partially observed data, while the original masked data are considered as ground truth. All training and inference processes are then conducted on the partially observed data, and those dropped values are kept for evaluation. Based on the partially observed data, we further conduct preprocessing methods, according to the recommended settings of each specific model. For our own SpaGAT and \modelname, we first normalize the total RNA counts of each cell, and then apply log1p transform. In addition, considering that the output format of the baseline models varies between raw counts and log-transformed values, we uniformly conduct postprocessing to make sure all imputed data and ground-truth data are normalized and log-transformed. By default, SpaGAT and \modelname adopt a bi-level masked autoencoder, while \modelname chooses a Cond PE as positional encoding. Reproducibility details and codes for all methods can be found on our GitHub repo~\footnote{Codes and hyperparameter settings are available at \url{https://github.com/wehos/CellT}} as well as Appendix~\ref{app:repro}. 

\noindent
\textbf{Evalutaion.} \label{sec:app-eval}
For imputation, the rooted mean squared error (RMSE), Pearson correlation coefficients (Pearson), and cosine similarity (Cosine) metrics are calculated based on the predictions of the masked values. For clustering, we first reduce the dimension of imputed data with PCA and then construct kNN graph based on the first $10$ PCs. The clustering result is obtained from Leiden algorithm on the kNN graph, where we conduct a grid search to find the optimal resolution for Leiden. Clustering metrics normalized mutual information (NMI) and adjusted Rand index (ARI) are then calculated based on the clustering results and predefined cell type labels. Lastly, all deep learning models are evaluated with $5$ random seeds, and the average performance is reported, while statistical models are evaluated only once. The standard deviation is presented in Appendix~\ref{app:std}.

\subsection{Imputation Performance}
In Table~\ref{tab:main}, we present the experimental results. It is shown that our \modelname consistently outperforms other baselines by showing a better imputation performance. Aside from that, there are several interesting observations. (1) scVI and DCA consistently present suboptimal performance, outperforming some advanced methods, e.g., GraphSCI. Since both DCA and scVI adopt ZINB-based autoencoders, this demonstrates the effectiveness of ZINB-based autoencoders. (2) Reference-based methods, i.e., gimVI, Tangram, SpaGE, generally do not work well on our highly noisy spatial transcriptomic dataset. These methods map spatial transcriptomic data to scRNA-seq reference datasets and impute the missing values based on reference data. However, the single-cell spatial transcriptomics data has a substantially low cover rate of RNA molecules, which causes distinct distributions between spatial transcriptomic data and reference scRNA-seq data. These results indicate the limitation of these reference-based imputation approaches. (3) SpaGAT, the non-transformer version of \modelname, presents fairly good performance. This demonstrates the advantage of our bi-level mask-autoencoder framework.  (4) Recent methods specifically designed for spatial transcriptomics either suffer from scalability issues or demonstrate suboptimal performance. Therefore, there remains significant space for exploration in the field of single-cell spatial transcriptomic imputation.

\subsection{Scalability Analysis.} Scalability is critical when deploying methods on single-cell spatial transcriptomic data since the datasets often contain tens of thousands of cells. We considered scalability as an important factor when conducting the evaluation. Specifically, we set limits on the runtime and memory consumption of the model, i.e., 48 hours of running time, 300G CPU memory, and 45G GPU memory. 
Notably, we encountered scalability issues in scGNN and Sprod methods even with the smallest Kidney dataset. scGNN and SEDR train neural networks on GPUs, however, both of them may run out of 300G CPU memory when building the cell-cell graph. Sprod separates data into numerous batches and purely runs on CPUs. We allocate 128 CPU processors for it, yet it fails to finish running in 48 hours. 
In contrast, our proposed \modelname exhibits \textbf{linear complexity w.r.t the number of cells}, thanks to our efficient autoencoder architecture and the Performer~\cite{choromanski2021rethinking} encoder. 

In addition, we present the empirical computational overhead of all methods on the largest Liver dataset in Table~\ref{tab:effi},. Specifically, we record the running time, peak CPU memory, and peak GPU memory. Peak CPU memory refers to the physical CPU memory consumption. Peak GPU Memory refers to the peak consumption of GPU memory resources. Running time refers to the training and inference for the imputation, excluding preprocessing time. The experiments are conducted on an RTX 8000 GPU card and 128 AMD EPYC CPU cores with 300G CPU memory. Each setting is repeated 4 times and the average running time is reported. Our \modelname method demonstrates superior computational efficiency, especially running time.



\begin{table}[ht]
\centering
\caption{ Computational overhead on Liver dataset. Methods are ranked by running time.}
\label{tab:effi}
\begin{tabular}{@{}cS[table-format=2.3]cc@{}}
\toprule
\textbf{Methods} & \parbox{1.5cm}{\centering \textbf{Running Time (hour)}} & \parbox{1.5cm}{\centering \textbf{Peak CPU Memory (Mb)}} & \parbox{1.5cm}{\centering \textbf{Peak GPU Memory (Mb)}} \\
\midrule
\modelname & \textbf{0.068} & 6,612 & 1,678 \\
scVI & 0.119 & \textbf{3,511} & 254 \\
Spage & 0.179 & 12,493 & {N/A} \\
DCA & 0.236 & 7,508 & 41,850 \\
Tangram & 0.551 & 4,507 & 17012 \\
gimVI & 0.971 & 6,600 & 250 \\
SAVER & 1.975 & 245,760 & {N/A} \\
scImpute & 2.917 & 269,490 & {N/A} \\
GraphSCI & 9.592 & 18,329 & {N/A} \\
\rowcolor{Gray} scGNN & {N/A} & {> 307,200} & {N/A} \\
\rowcolor{Gray} SPROD & {>48} & {N/A} & {N/A} \\
\rowcolor{Gray} SEDR & {N/A} & {>307,200} & {N/A} \\
\bottomrule
\end{tabular}
\vspace{-2em}
\end{table}


\begin{figure}[ht]     \centering
    \caption{Clustering performance on imputed data of Lung dataset.} 
     \includegraphics[width=0.9\linewidth]{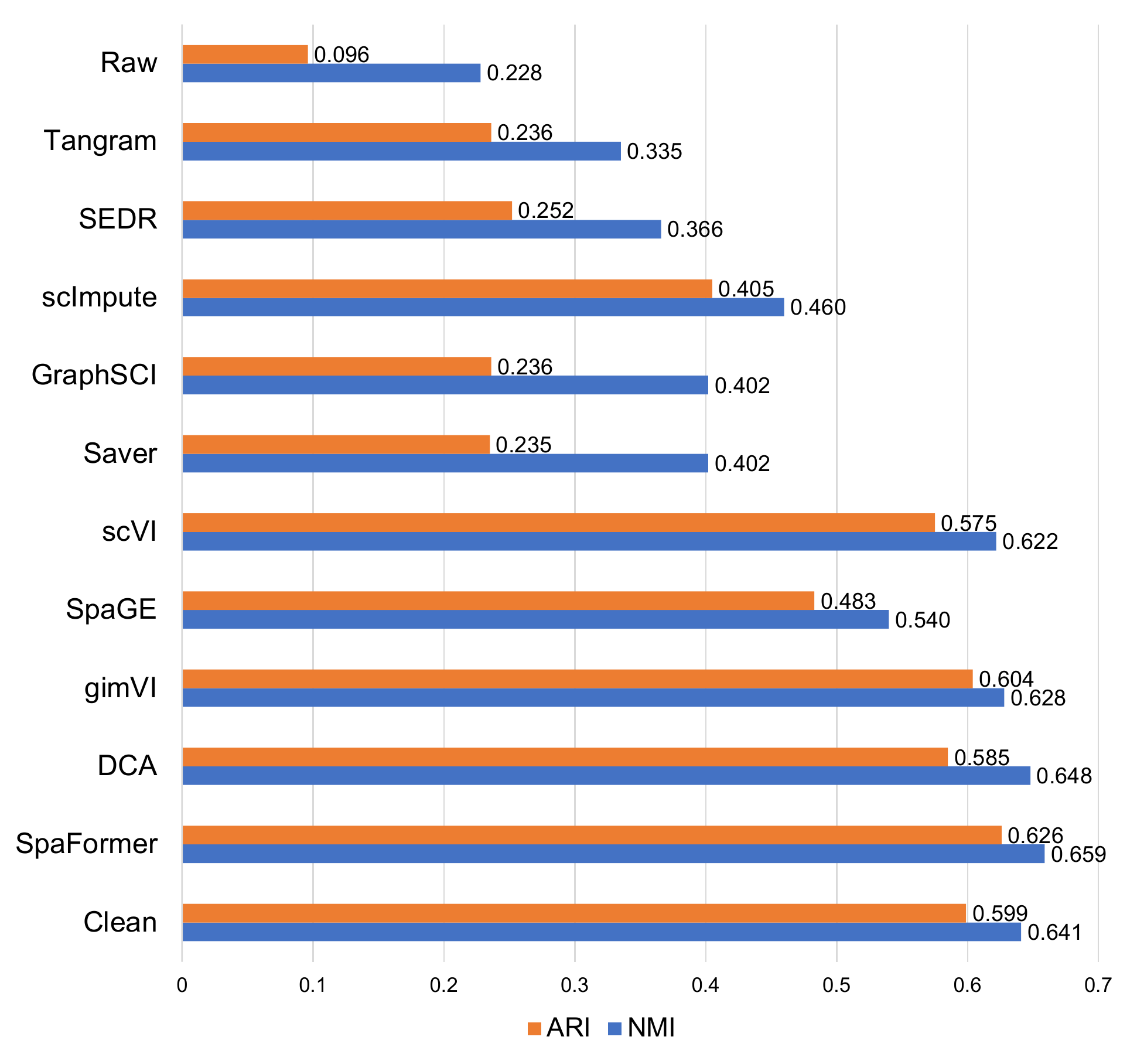}
    \vskip -1em

\label{fig:cls}
\end{figure}

\subsection{Clustering Performance}
In addition to imputation performance, we conduct unsupervised clustering on the imputed data to validate whether the imputation can help recover cell type information. For evaluation, cell type labels accompanied by the dataset are considered as ground truth. It is expected that well-imputed data can better recover the cell clustering structures. As shown in Figure~\ref{fig:cls}, although all models enhance the clustering performance as compared to the unimputed raw data, clustering based on \modelname achieves the best performance. 

In addition, an interesting observation is that despite reference-based methods (i.e., SpaGE and gimVI) do not achieve impressive performance on imputation metrics, their clustering performance is relatively better. This indicates that they may have succeeded in discovering reference cells from scRNA-seq data, however, they still suffer from the distribution gap between the reference and query data.

\subsection{Ablation Study}
\subsubsection{Positional Encodings}
\begin{figure}[ht]
     \centering
     \vskip -0.5em

\includegraphics[width=0.95\linewidth]{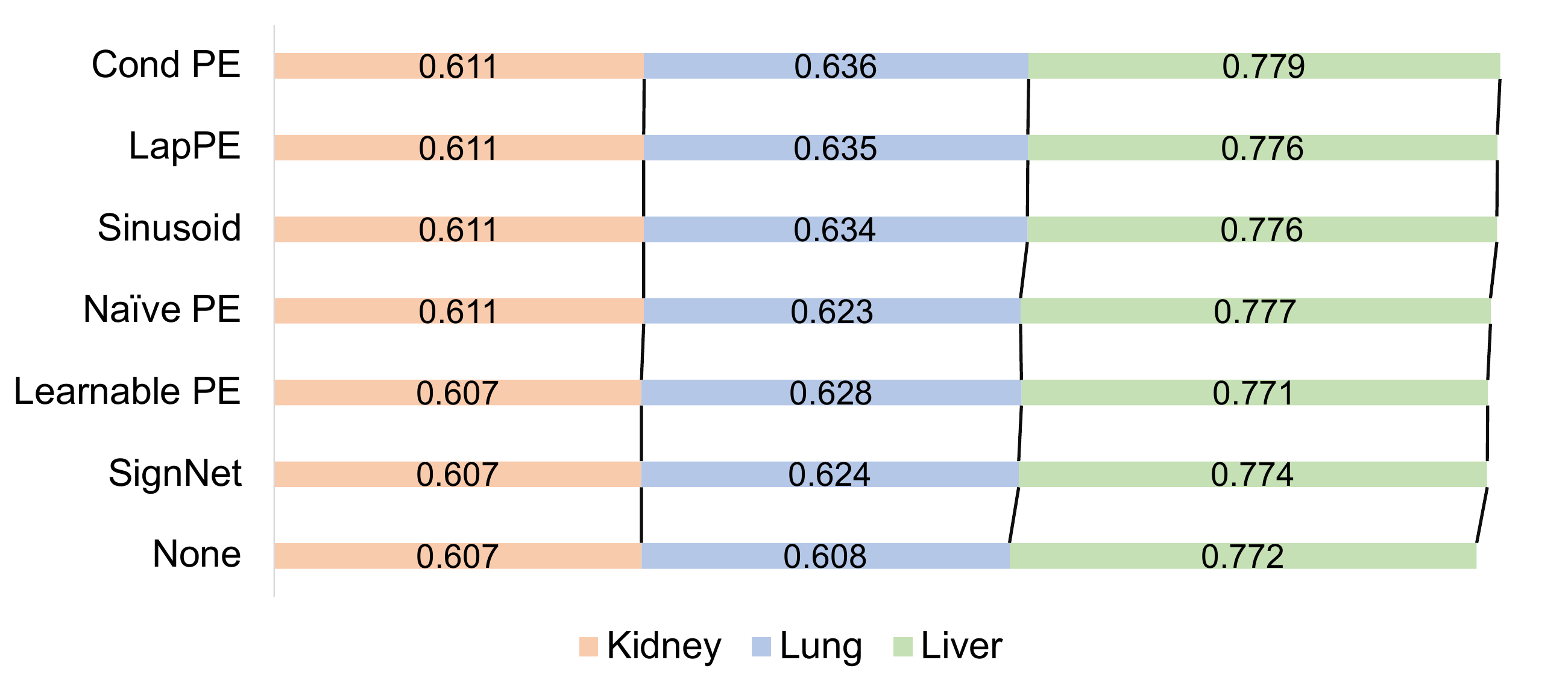}
\vspace{-1em}

     \caption{Comparison between different positional encodings on three datasets. Values indicate Pearson correlation coefficient.} \label{fig:pe}
     \vskip -0.5em
     
\end{figure}

As we mentioned in Section~\ref{sec:pe}, we conduct experiments on three datasets to compare the performance of different positional encodings. According to our experiments in FIgure~\ref{fig:pe}, Cond PE and LapPE outperform other positional encodings w.r.t overall performance on three datasets, indicating the effectiveness of graph-based PEs. 
Meanwhile, Cond PE outperforms SignNet, suggesting that global effectiveness might not be a desired property for spatial transcriptomics, as we discussed in Section~\ref{sec:pe}. In conclusion, we select Cond PE as a default setting in \modelname. Besides, a 2-dimensional Sinusoid PE can also be a good choice for spatial transcriptomics since it has fewer parameters than others. Meanwhile, the gap between transformers with and without positional encodings is less significant than we expected. Therefore, there is still large room for exploration in positional encodings for spatial transcriptomic data.

\subsubsection{Autoencoder Frameworks}
\begin{figure}[ht]
     \centering
\includegraphics[width=0.85\linewidth]{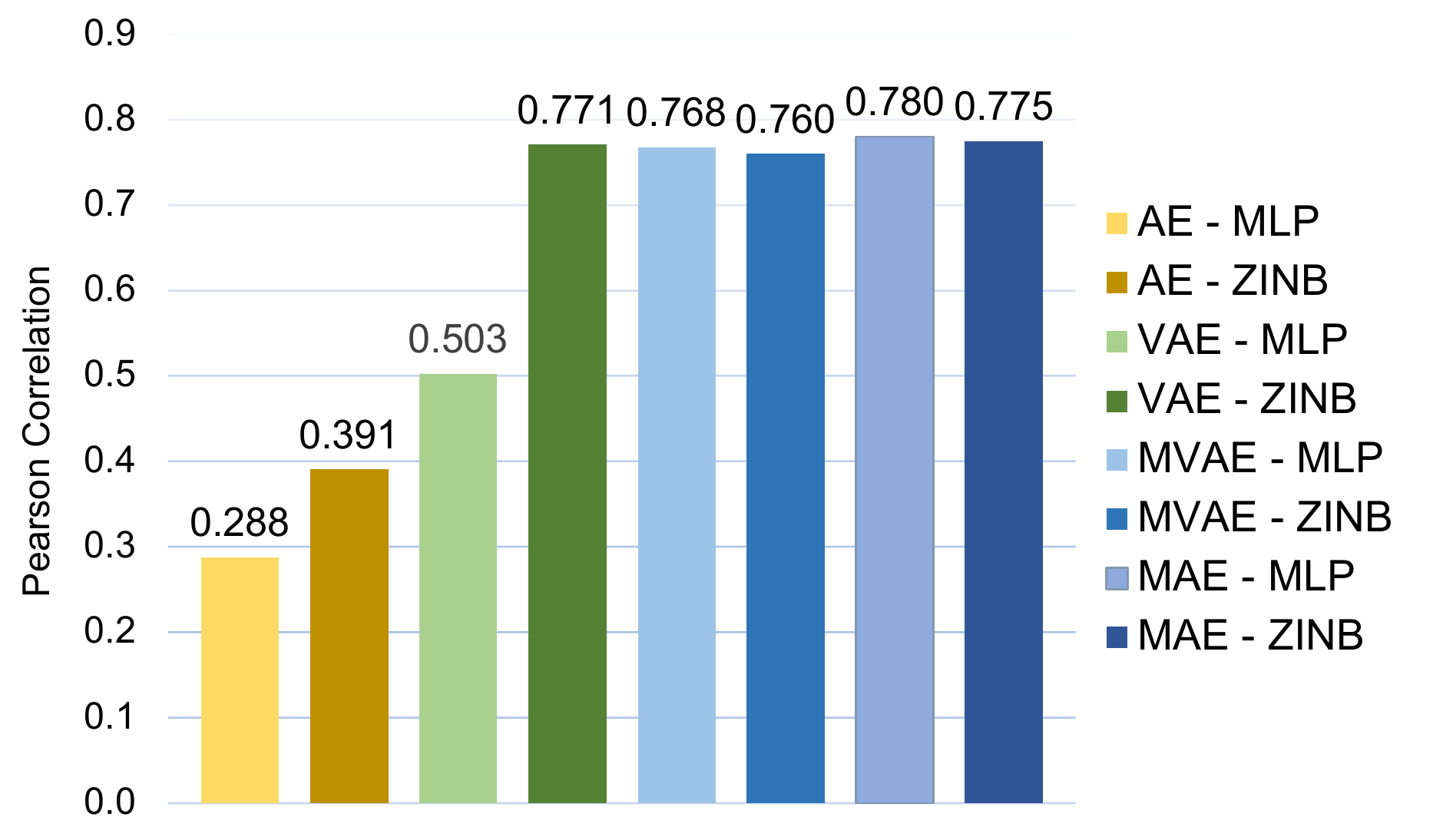}
       \vspace{-0.5em}
    \caption{Ablation study on different autoencoder variants, i.e., vanilla autoencoders (AE), variational autoencoders (VAE), bi-level masked autoencoder (MVAE) and masked autoencoder (MAE). For each variant, we implement both MLP and ZINB decoders. Values indicate the Pearson correlation coefficient on the Liver dataset.} \label{fig:ae}
    \vspace{-2.5em}
\end{figure}

Another highlight of our \modelname framework is that we generalize different popular autoencoder-based models. Based on our general framework, we conduct a thorough ablation study on the Liver dataset. The experimental results in Figure~\ref{fig:ae} indicate that various variations of autoencoders achieved optimal performance when combined with masking. Additionally, the simple bi-level mask autoencoder obtained the best results. Therefore, we recommend the masked autoencoder as the default setting for our \modelname framework.

\begin{figure}[ht]

     \centering
     \includegraphics[width=0.6\linewidth]{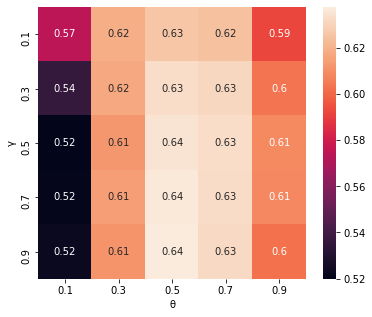}
     \vspace{-1em}
     \caption{Parameter analysis on $\theta$ and $\gamma$. Values indicate Pearson correlation coefficient on Lung dataset.} \label{fig:msk}
     \vspace{-1em}

\end{figure}

\subsubsection{Mask Ratio}

Lastly, we conduct a parameter analysis on token masked ratio $\theta$ and feature masked ratio $\gamma$ in Section~\ref{sec:mae} to demonstrate the importance of our bi-level masking strategy. As shown in Figure~\ref{fig:msk}, joint tuning $\theta$ and $\gamma$ results in an optimal Pearson correlation score. In our main experiment, we select $\theta=0.5$, $\gamma=0.5$ as default parameters, since this setting consistently achieves optimal performance on three datasets.

\section{Related Works}
The increased resolution of transcriptomics profiling methods comes at the cost of increasing data sparsity. The profiling technology may miss a substantial number of expressed genes of an individual cell due to a low mRNA capture rate.  
A popular way to address this issue is to perform imputation, which aims to correct false zeros by estimating realistic values for those gene-cell pairs. 
A large number of methods have been developed for the task of scRNA-seq data imputation, mainly focusing on generative probability models or matrix factorization~\cite{ gong2018drimpute, huang2018saver,  ronen2018netsmooth, van2017magic}. Aside from these methods, deep learning models have gained immense popularity over recent years. A natural deep learning architecture for the imputation task is the autoencoder, due to its prevalence in data denoising and missing data applications~\cite{beaulieu2017missing, boquet2019missing, gondara2017multiple, gondara2018mida, pereira2020reviewing}. To leverage the spatial information of spatial transcriptomic data, recent methods Sprod~\cite{wang2022sprod} and SeDR~\cite{xu2024unsupervised} construct graphs that connect cells within local neighborhood. However, these local graphs hinder the identification of cells with long-range correlations. Different from these methods, we propose to utilize transformers for spatial transcriptomic imputation to globally model the cell-cell interactions while exploiting the spatial information. 

\section{Conclusion}

In this work, we comprehensively investigate two key questions for transformer-based spatial transcriptomic imputation at the cellular level: (1) \textit{how to encode spatial information of cells in transformers}, and (2) \textit{ how to train a transformer for spatial transcriptomic imputation}. By answering these two questions, we proposed a transformer-based general imputation framework \modelname for single-cell spatial transcriptomics data. In addition, we propose a new bi-level masking technique, which can be incorporated into general autoencoder frameworks. Our method outperforms all existing methods for recovering missing values and clustering structures while demonstrating superior computational efficiency.   

\vspace{-0.5em}



\bibliographystyle{ACM-Reference-Format}
\bibliography{main}

\appendix
\section{Positional Encodings} \label{app:pe}
\subsection{Learnable PE}
Learnable PE is introduced in ViT~\cite{dosovitskiy2020image}, where it independently learns a positional encoding for each individual position. 
This is equivalent to creating a learnable lookup table $\mathbf{P} \in \mathcal{R}^{100 \times 100 \times d}$, where $\mathbf{P}_{x,y,:}$ is the $d$ dimensional positional encoding for patch $(x, y)$.

\subsection{Sinusoid PE}
Sinusoid PE is proposed in vanilla transformer~\cite{vaswani2017attention}. The original version is designed for sequential data, while we can extend it to 2-dimensional space similar to DETR~\cite{carion2020end} and MI2LaTeX~\cite{wang2021translating}. 
Specifically, we formulate the 2-dimensional sinusoid PE as:
\begin{equation}
\begin{aligned} \label{eq:sinu1}
& \operatorname{PE}(x, y, 2 i)=\sin \left(x / 10000^{4 i / d}\right), \; \operatorname{PE}(x, y, 2 i+1)=\cos \left(x / 10000^{4 i / d}\right), \\ 
& \operatorname{PE}(x, y, 2 j+d / 2)=\sin \left(y / 10000^{4 j / d}\right), \; \operatorname{PE}(x, y, 2 j+1+d / 2)=\cos \left(y / 10000^{4 j / d}\right), 
\end{aligned}
\end{equation}
where $i,j \in [0, d/4)$ specify the feature dimensions. $(x, y)$ denotes 2D coordinates.

\subsection{Naive PE}
naive PE projects spatial coordinates $\widetilde{\mathbf{C}}$ to desired dimension $d$ via a linear transformation or MLP. One example is $\mathbf{P} = \sigma\left( \widetilde{\mathbf{C}} \mathbf{W} + b \right)$
where $\mathbf{W} \in \mathcal{R}^{2 \times d}$ and $b \in \mathcal{R}^{d}$ are learnable parameters, $\sigma$ is a non-linear function, and $\mathbf{P} \in \mathcal{R}^{n \times d}$ is the positional encodings. 

\subsection{Random Walk PE}
Random walk PE is proposed in LSPE~\cite{dwivedi2021graph}, defined as:
\begin{displaymath}
    \mathbf{p}_i^{\mathrm{RWPE}}=\left[\mathrm{RW}_{i i}, \mathrm{RW}_{i i}^2, \cdots, \mathrm{RW}_{i i}^k\right] \in \mathbb{R}^k,
\end{displaymath}
where $\mathrm{RW}=\mathbf{A} \mathbf{D}^{-1}$ is the random walk operator, $\mathbf{D}$ is the degree matrix, and $k$ is the number of random walk steps. 

\subsection{Summary} \label{sec:app-dis}
\noindent\textbf{Distance awareness} refers to the property that a positional encoding is able to indicate the pair-wise distance between two tokens, which is the most fundamental property that is needed by transformers for spatial transcriptomics, as mentioned in Section~\ref{sec:intro}. The only encoding without this feature is RWPE. This is because RWPE aims for encoding local graph structure instead of relative positions.

\noindent\textbf{Global effectiveness} refers to the property that a positional encoding is able to indicate the relation between distant tokens. In other words, even if two tokens are very far apart, the positional encoding can still measure their distance. This is a double-edged property. On the one hand, global PE enhances the ability of the transformer to capture long-range relevance between cells.  On the other hand, it weakens the inductive bias of locality. This inductive bias, however, is effective for local cell-cell interactions. Cond PE imposes locality on the spatial adjacency graph, distinct from most other PEs.

\noindent\textbf{Translation invariance} refers to the property that a positional encoding stays invariant when the overall spatial coordinates are translated. Many previous studies~\cite{liu2021swin, shaw2018self} have shown that this property is effective in enhancing the generalization ability of transformers. This property is also effective for spatial transcriptomics, and we therefore consider it an important feature in the selection of PE. Fortunately, graph-based position encodings typically possess this desirable property, as the translation of coordinates does not affect relative positions and therefore does not alter the spatial adjacency graph.

\noindent\textbf{Structure awareness} refers to the property that a positional encoding is able to encode the structure of space, such as density and homogeneity within local space, and the global spatial structure. This ability often comes from encoding the spatial graph, which goes beyond the definition of position encodings, but may provide additional information for transformers. 

\section{Reproduciblity Details} \label{app:repro}

\subsection{Data Availability}
Two datasets (Lung and Liver) we used are public available on Nanostring official website: \url{https://nanostring.com/products/cosmx-spatial-molecular-imager/nsclc-ffpe-dataset/}.

\subsection{Code Availability}
Our codes for reproducibility will be released on GitHub.

\subsection{Implementation Settings}
\subsubsection{Baselines}
To evaluate the effectiveness of \modelname{}, we compare it with the state-of-the-art spatial and non-spatial transcriptomic imputation models:
(1) {scImpute}~\cite{li2018accurate} employs a probabilistic model to detect dropouts, and implements imputation through non-negative least squares regression.
(2) {SAVER}~\cite{huang2018saver} uses negative binomial distribution to model the data and estimates a Gamma prior through Poisson Lasso regression. The posterior mean is used to output expression with uncertainty quantification from the posterior distribution.
(3) {scVI}~\cite{lopez2018deep} models dropouts with a ZINB distribution, and estimates the distributional parameters of each gene in each cell with a VAE model.
(4) {DCA}~\cite{eraslan2019single} is an autoencoder that predicts parameters of chosen distributions like ZINB to generate the imputed data. 
(5) {GraphSCI}~\cite{rao2021imputing} employs a graph autoencoder on a gene correlation graph. Meanwhile, it uses another autoencoder to reconstruct the gene expressions, taking graph autoencoder embeddings as additional input.
(6) {scGNN}~\cite{wang2021scgnn} first builds a cell-cell graph based on gene expression similarity and then utilizes a graph autoencoder together with a standard autoencoder to refine graph structures and cell representation. Lastly, an imputation autoencoder is trained with a graph Laplacian smoothing term added to the reconstruction loss.
(7) {gimVI}~\cite{lopez2019joint} is a deep generative model for integrating spatial transcriptomics data and scRNA-seq data which can be used to impute spatial transcriptomic data. gimVI is based on scVI~\cite{lopez2018deep} and employs alternative conditional distributions to address technology-specific covariate shift more effectively.
(8) {Sprod}~\cite{li2022cell} is the latest published imputation method for spatial transcriptomics data. It first projects gene expressions to a latent space, then connects neighboring cells in the latent space, and prunes it with physical distance. Then a denoised matrix is learned by jointly minimizing the reconstruction error and a graph Laplacian smoothing term. 
(9) {SpaGAT} is a baseline model created by ourselves. It is the same bi-level masking autoencoder framework as \modelname, based on a graph neural network encoder with spatial graphs. Specifically, we implement a graph attention network~\cite{velivckovic2017graph} as the encoder. Since the graph attention network is a localized version of transformers, SpaGAT can be considered an ablation study for our \modelname model.

\subsubsection{Hyperparameter Settings}
For our own2 SpaGAT and SpaFormer, we first normalize the total RNA counts of each cell, and then apply log1p transform. We heavily conducted hyperparameters searching on the Lung dataset. However, we noticed that the performance is not sensitive to most hyperparameters, except for masking rate, autoencoder type, and positional encodings as we presented in ablation studies. To reproduce our results, the recommended hyperparameters are n\_layer=2, num\_heads=8, num\_hidden=128, latent\_dim=20, learning rate=1e-3, weight\_decay=5e-4. We used these default hyperparameters in the other two datasets. The source codes will been released on our github. For our own created baseline model SpaGAT, we used the same set of hyperparameters while replacing the transformer encoder with a graph attention network.

For baseline models, all the implementations are from the authors’ repo/software. Optimizers/trainers are provided by original implementations. Preprocessings are also consistent with the original methods. The detailed settings are as below:

ScImpute only involves two parameters. Parameter K denotes the potential number of cell populations, threshold t on dropout probabilities. We set t=0.5 and K=15. This is per the author’s instructions in their paper, i.e., a default threshold value of 0.5 is sufficient for most scRNA-seq data, and K should be chosen based on the clustering result of the raw data and the resolution level desired by the users, where K=15 is close to the ground-truth cell-type number.

SAVER is a statistical model and does not expose any hyperparameters in their implementation. Therefore, we run their default setting.

For scVI, we searched for combinations of n\_hidden=[128, 256], n\_layer=[1, 2], gene\_likelihood=[nb, zinb] on the Lung5 dataset. All settings are repeated five times and the best mean performance is achieved by n\_hidden=128, n\_layer=1, and gene\_likelihood=’nb’. Other parameters are per default. We then applied this set of hyperparameters to all three datasets and reported it in the main results.

For DCA, the original hyperparameter optimization mode is broken (there is a relevant unresolved issue on GitHub), and there are no other parameter instructions in the tutorial, therefore we used default parameters.

For GraphSCI, we tried hidden1=[16, 32, 64], hidden2=[32, 64]. Note that GraphSCI does not have other hyperparameters and the default number of hidden sizes is quite small. The performance reported in our paper was obtained from hidden1=32, and hidden2=64.

gimVI software does not expose hyperparameters such as n\_hidden and n\_layer, so we follow the default settings. Additionally, gimVI requires external scRNA-seq reference data. For the Lung5 dataset, we used data from GSE131907 [8] as a reference. For Kidney, we used data from GSE159115 [9], and for Liver we used data from GSE115469 [10].

scGNN provides two sets of preprocessing in their GitHub repo and we adopt the first one "Cell/Gene filtering without inferring LTMG" for CSV format. Then we follow the corresponding instructions to run their imputation method but get an out-of-memory error during EM algorithm.

Sprod provides batch mode for handling very big spatial datasets. We follow their instructions for dataset without a matching image and ran the batch mode with num\_of\_batches=30, however, it can not finish running within 48 hours even on the smallest Kidney dataset.

For Spage we follow the official tutorial. For SeDR, we follow their official tutorial for imputation and batch integration, where we consider each FOV as a batch. SeDR does not provide any information on hyperparameters in their tutorial, therefore, we run their default method with 5 random seeds. To adapt SeDR's output to our evaluation protocol, we remove the inherent standardization function from SeDR's `.recon` function.

For Tangram, we found that directly inputting all the data would lead to OOM issue. Therefore, we follow the official instruction to separate the data into 5 splits and impute them split by split.

\section{Standard Deviation} \label{app:std}

{\tiny
\begin{table*}[h]
\centering
\begin{tabular}{|c|c|c|c|c|c|c|c|c|c|}
\hline & Lung 5 & Lung 5 & Lung 5 & Kidney 1139 & Kidney 1139 & Kidney 1139 & Liver Norma & Liver Normal & Liver Norma \\
\hline & RMSE & Pearson & Cosine & RMSE & Pearson & Cosine & RMSE & Pearson & Cosine \\
\hline scVI & $\begin{array}{l}0.2861\pm \\
0.0037\end{array}$ & $\begin{array}{l}0.6231\pm \\
0.0136\end{array}$ & $\begin{array}{l}0.6661\pm \\
0.0112\end{array}$ & $\begin{array}{l}0.2901\pm \\
0.0040\end{array}$ & $\begin{array}{l}0.5834\pm \\
0.0152\end{array}$ & $\begin{array}{l}0.6480\pm \\
0.0125\end{array}$ & $\begin{array}{l}0.2797\pm \\
0.0159\end{array}$ & $\begin{array}{l}0.5749\pm \\
0.0641\end{array}$ & $\begin{array}{l}0.6224\pm \\
0.0582\end{array}$ \\
\hline DCA & $\begin{array}{l}0.2858\pm \\
0.0002\end{array}$ & $\begin{array}{l}0.6223\pm \\
0.0008\end{array}$ & $\begin{array}{l}0.6648\pm \\
0.0007\end{array}$ & $\begin{array}{l}0.2852\pm \\
0.0005\end{array}$ & $\begin{array}{l}0.5985\pm \\
0.0022\end{array}$ & $\begin{array}{l}0.6597\pm \\
0.0017\end{array}$ & $\begin{array}{l}0.2542\pm \\
0.0129\end{array}$ & $\begin{array}{l}0.657\pm \\
0.0461\end{array}$ & $\begin{array}{l}0.688\pm \\
0.0401\end{array}$ \\
\hline GraphSCI & $\begin{array}{l}0.3957\pm \\
0.0006\end{array}$ & $\begin{array}{l}0.1334\pm \\
0.0016\end{array}$ & $\begin{array}{l}0.3081\pm \\
0.0007\end{array}$ & $\begin{array}{l}0.3624\pm \\
0.0009\end{array}$ & $\begin{array}{l}0.2403\pm \\
0.0048\end{array}$ & $\begin{array}{l}0.4128\pm \\
0.0029\end{array}$ & $\begin{array}{l}0.3347\pm \\
0.0009\end{array}$ & $\begin{array}{l}0.3707\pm \\
0.0065\end{array}$ & $\begin{array}{l}0.4430\pm \\
0.0065\end{array}$ \\
\hline gimVI & $\begin{array}{l}0.3170\pm \\
0.0018\end{array}$ & $\begin{array}{l}0.5320\pm \\
0.0063\end{array}$ & $\begin{array}{l}0.5917\pm \\
0.0052\end{array}$ & $\begin{array}{l}0.4387\pm \\
0.0015\end{array}$ & $\begin{array}{l}-0.0104\pm \\
0.0004\end{array}$ & $\begin{array}{l}0.1967\pm \\
0.0027\end{array}$ & $\begin{array}{l}0.4542\pm \\
0.0024\end{array}$ & $\begin{array}{l}-0.0015\pm \\
0.0010\end{array}$ & $\begin{array}{l}0.1167\pm \\
0.0050\end{array}$ \\
\hline SpaFormer & $\begin{array}{l}0.2785\pm \\
0.0005\end{array}$ & $\begin{array}{l}0.6363\pm \\
0.0019\end{array}$ & $\begin{array}{l}0.6786\pm \\
0.0016\end{array}$ & $\begin{array}{l}0.2794\pm \\
0.0042\end{array}$ & $\begin{array}{l}0.6108\pm \\
0.0152\end{array}$ & $\begin{array}{l}0.671\pm \\
0.0121\end{array}$ & $\begin{array}{l}0.2117\pm \\
0.0014\end{array}$ & $\begin{array}{l}0.7976\pm \\
0.0034\end{array}$ & $\begin{array}{l}0.7797\pm \\
0.0032\end{array}$ \\
\hline
\end{tabular}
\end{table*}
}

\end{document}